\documentclass{bmvc2k}

\usepackage{times}
\usepackage{epsfig}
\usepackage{graphicx}
\usepackage{amsmath}
\usepackage{amssymb}
\usepackage{dsfont}
\usepackage{nicefrac}
\usepackage{arydshln}
\usepackage{xspace}
\usepackage{rotating}
\usepackage{caption}
\usepackage[symbol]{footmisc}

\usepackage{booktabs}

\DeclareMathOperator*{\argmin}{arg\,min}

\newcommand{\model}[1]{{\fontfamily{qpl}\selectfont {\scriptsize #1}}}
\newcommand{\dataset}[1]{{\fontfamily{phv}\selectfont {\scriptsize #1}}}
\newcommand{\tableHeading}[1]{{\fontfamily{lmr}\selectfont {\scriptsize #1}}}
\newcommand{\myparagraph}[1]{\vspace{0.5pt}\noindent{\bf #1}}

\title{Conditional De-Identification of 3D Magnetic Resonance Images}

\addauthor{Lennart Alexander Van der Goten}{lavdg@kth.se}{1,2}
\addauthor{Tobias Hepp}{tobias.hepp@tuebingen.mpg.de}{3,4}
\addauthor{Zeynep Akata}{zeynep.akata@uni-tuebingen.de}{3,4}
\addauthor{Kevin Smith}{ksmith@kth.se}{1,2}
\addauthor{}{}{}

\addinstitution{
 KTH Royal Institute of Technology\\
 Stockholm, SWEDEN
}
\addinstitution{
 Science for Life Laboratory\\
 Solna, SWEDEN
}
\addinstitution{
 Max Planck Institute for Intelligent Systems\\
 Tübingen, GERMANY
}
\addinstitution{
 University of Tübingen\\
 Tübingen, GERMANY
}
\runninghead{VAN DER GOTEN ET AL.}{CONDITIONAL DE-IDENTIFICATION OF MRI SCANS}

\def\eg{\emph{e.g}\bmvaOneDot}
\def\ie{\emph{i.e}\bmvaOneDot}

\def\etal{\emph{et al}\bmvaOneDot}

\begin{document}

\maketitle

\begin{abstract}
Privacy protection of medical image data is challenging.
Even if metadata is removed, brain scans are vulnerable to attacks that match renderings of the face to facial image databases.
Solutions have been developed to de-identify diagnostic scans by obfuscating or removing parts of the face.
However, these solutions either fail to reliably hide the patient's identity or are so aggressive that they impair further analyses.
We propose a new class of de-identification techniques that, instead of removing facial features, remodels them. 
Our solution relies on a conditional multi-scale GAN architecture.
It takes a patient's MRI scan as input and generates a 3D volume conditioned on the patient's brain, which is preserved exactly, but where the face has been de-identified through remodeling.
We demonstrate that our approach preserves privacy far better than existing techniques, without compromising downstream medical analyses. Analyses were run on the OASIS-3 and ADNI corpora. 

\end{abstract}

\section{Introduction}%
\begin{figure}[t]
\begin{center}
\begin{tabular}[t]{@{}c@{}}
\includegraphics[width=1\columnwidth]{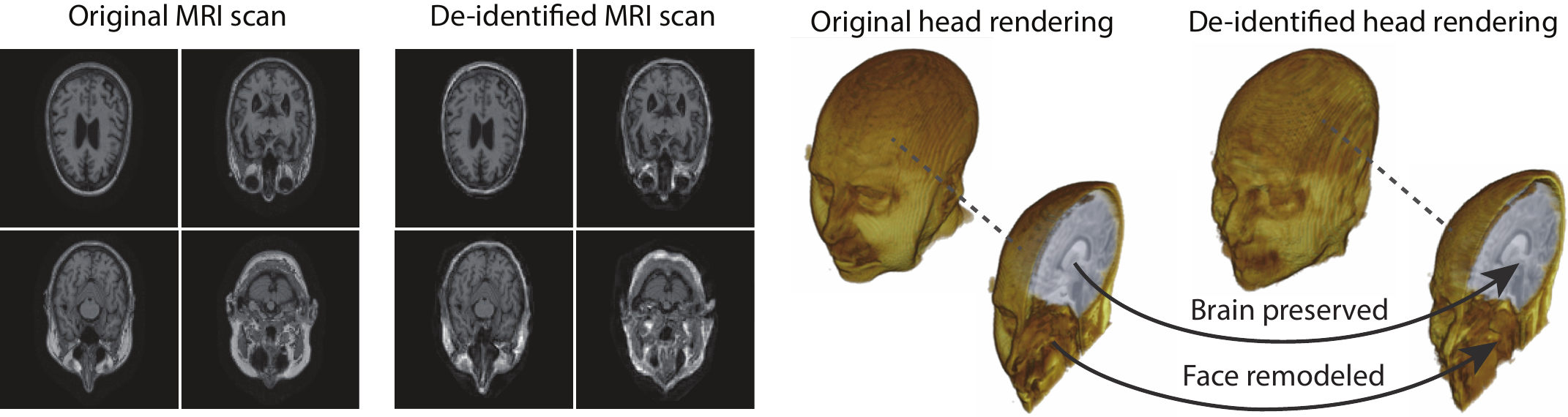} \\
\end{tabular}
\vspace{-2mm}
\caption{\emph{De-identification through facial remodeling.} Face renderings from medical scans such as these MRIs represent a privacy vulnerability.
To address this, we define a new class of de-identification techniques that aims to realistically \emph{remodel} privacy-sensitive regions, such as the face, while \emph{preserving} essential regions, such as the brain (illustrated \dataset{OASIS-3}). 
}
\label{fig:de-identification}
\vspace{-5mm}
\end{center}
\end{figure}

The digitalization of heath records has increased the risk of --and impact of-- large scale data leaks.
Although data compliance standards have been enacted to protect health records (HIPAA and GDPR), privacy of medical data is a growing concern.
Three-dimensional scans such as magnetic resonance images (MRI) and computed tomography (CT), for example, contain an intrinsic privacy risk~\cite{lotan2020medical}.
Detailed renderings of the head can be crafted from 3D scans using techniques such as volumetric raycasting, as in Figure~\ref{fig:de-identification}.
This vulnerability can expose the patient's identity if the renderings are matched to a face database~\cite{mazura2012facial,lotan2020medical}.

To prevent these types of attack, medical scans are currently de-identified using crude \textit{removal-based} techniques~\cite{Bischoff-Grethe2007,Schimke2011,Milchenko2013} which seek to remove privacy-sensitive parts of the head (examples in Figure \ref{fig:qualitative results}).
However, as we demonstrate, these existing techniques fail to reliably hide the patient's identity -- or they are so aggressive that they impair further medical analyses.
A better solution is needed.

One might ask \emph{why de-identify the face when one can just remove everything except the brain?}
This approach, known as skull-stripping~\cite{Segonne2004}, does provide excellent privacy guarantees, but unfortunately \textbf{renders the scan useless for many types of clinical analysis}.
Automated tools for analyzing MRI scans rely on landmarks within the head, and fail when they are removed~\cite{DeSitter2020}.
Furthermore, skull-stripping corrupts measurements of important tissues and fluids, such as extra-cranial CSF~\cite{Bischoff-Grethe2007}.
For these reasons, \emph{remodeling the head} rather than deleting privacy-sensitive regions is desirable, because it protects privacy and at the same time ensures robustness of downstream medical analyses.

Therefore, in this work,  we define a new class of de-identification techniques that \emph{remodels} the privacy-sensitive regions without altering the content of medically relevant data (see Figure~\ref{fig:de-identification}). 
Under such a remodeling approach, the face, eyes, oral and nasal cavities, \emph{etc.}~should exhibit realistic appearance and structure of appropriate size, but should otherwise be independent of the original data.
To solve this task, we propose a novel model called \emph{Convex Privacy} GAN, or \model{CP-GAN}, that conditions on a convex hull of the skull extracted from the scan to be de-identified.
The generator learns to synthesize volumes that \textit{preserve} medically-sensitive regions such as the brain, while non-invertibly \emph{remodeling} privacy-sensitive characteristics from the original scan.

The main contributions of this work are as follows:
    (1) We define a novel methodology to ensure privacy in medical imagery in which medically relevant regions are preserved and privacy-sensitive regions are de-identified.
    (2) We propose \model{CP-GAN}, a conditional multi-scale volumetric GAN that realizes a solution to the aforementioned methodology.
    (3) Through human- and model-based experiments, we show that \model{CP-GAN} preserves privacy in MRI scans more reliably than removal-based techniques without adversely affecting downstream analyses.
In addition, we make technical contributions towards the generation of the convex hull and surface representations necessary for the privacy conditioning of the GAN.
Source code as well as a video demonstration can be found in the \emph{supplementary material}.

\section{Related Work}
\label{sec:related}

A handful of de-identification techniques exist for MRI scans, which are conventionally used for sharing and distribution of MRI data.
These existing methods rely on a \emph{removal} approach to privacy. 
\model{DEFACE} \cite{Bischoff-Grethe2007} estimates the probabilities of voxels belonging to the face based on an atlas of healthy control subjects. 
The scan is de-identified by setting intensities of voxels whose probabilities are small enough to zero. 
\model{QUICKSHEAR} \cite{Schimke2011} is a fast but simple approach that computes a hyperplane to divide the MRI into two regions: one containing facial structures, and the other containing the brain of the scan. 
Voxels in the first part are set to zero. 
\model{FACE MASK} \cite{Milchenko2013} uses a filtering method to blur the facial features. 
These existing de-identification approaches are based on traditional computer vision techniques; we believe that the proposed algorithm is the first to adopt a learning-based approach.

While not a de-identification method, \emph{Shin} \etal ~\cite{Shin2018} recently proposed a \emph{pix2pix}-inspired  model~\cite{Isola2016} to generate synthetic abnormal MRI images with brain tumors.
In this work, the authors argue that, in principle, their approach can be used to generate a completely artificial corpus where none of the scans can be attributed to actual patients.
However, as the brain data is hallucinated, this method is not useful for our task.

The literature covering removal of privacy-sensitive information from image data largely focuses on de-identification of photographs of faces~\cite{jourabloo2015attribute,newton2005preserving}.
Among these, \emph{Deep Privacy}~\cite{Hukkelas2019} is the closest to our approach  as it was the first to suggest GANs to de-identify faces.
It conditions on an \emph{a priori} binary segmentation, guiding the generator to inpaint privacy-sensitive regions while preserving insensitive regions.
Similar to our approach, \emph{Deep Privacy} seeks to anonymize faces -- but in 2D images. 
To identify face regions for conditional inpainting, it relies on an SSD detector~\cite{Liu2015}.
We develop an alternative approach because a 3D analogue does not exist, and this allows us to comprehensively remodel interior regions of the head such as the neck and oral/nasal cavities.
In particular, we define a convex hull enclosing the head and mask of the brain for conditioning.
Finally, whereas \emph{Deep Privacy} de-identifies conventional images of size $128{\times}128$, our goal is to generate  much higher dimensional 3D volumes at $128^3$ voxels -- the equivalent of a $1448{\times}1448$ image.

  \section{Conditional De-Identification of 3D Images}
 \label{sec:method}

Given a set of 3D images $ (X^{(i)})_{i=1,\ldots N} \overset{\text{i.i.d.}}{\sim} \mathcal{P}_{X}$ with values in $\mathcal{I}^{S \times S \times S}$ over some intensity space $\mathcal{I} \subset \mathbb{R}$, 
we are interested in finding a function of the form $Y{=} f_\Phi(\gamma(X)) {\sim} \mathcal{P}_{Y}$ that maps a 3D image $X$ to its de-identified counterpart $Y$.
The task of the function $\gamma(X)$ is to filter out any sensitive information in order to make it impossible to infer the subject's identity given only $Y$; \ie to create a \emph{privacy preserving representation}.
In this work we consider MRI data\footnote{MRI scans can be acquired under different conditions, data availability limits us to common T1-weighted MRI.}, and choose $\gamma(X)$ to be a function of the convex hull of the head $c(X) {\in} \lbrace 0, 1 \rbrace^{S \times S \times S}$ and the brain mask $b(X) {\in} \lbrace 0, 1 \rbrace^{S \times S \times S}$. 

Within this \emph{remodeling}-based privacy framework, we impose three requirements: (i) \emph{Anonymity}, \ie $\gamma(X)$ is non-invertible; (ii) \emph{Distribution preservation}, \ie $\mathcal{P}_{X}$ and $\mathcal{P}_{Y}$ are stochastically indistinguishable and finally (iii)  \emph{Brain preservation}, \ie $\forall (i, j, k): b(X)_{i, j, k} = 1 \implies X_{i, j, k} = f_\Phi(\gamma(X))_{i, j, k}$.
In other words, we are interested in deriving a function $f_\Phi$ that maps some original scan $X$ to some de-identified scan $Y$, while retaining medically relevant information (\eg the brain) but preventing other information specific to $X$ to leak into $Y$ (\eg the face).
This makes it impossible to infer a person's identity from facial renderings. 
Figure \ref{fig:overview} depicts the de-identification process, described below, including the privacy transform $\gamma(X)$ and the mapping function $f_\Phi$ implemented with a conditional multi-scale volumetric GAN.

 \subsection{The Privacy Transform $\gamma(X)$}
 \label{sec:preprocessing}
The goal of the privacy transform is to non-invertibly change an individual MRI representation $x$ into a form $\gamma(x)$ that removes detailed privacy-sensitive information and replaces it with a convex hull filled with 1's, smoothing away detailed face information (\eg eyes, nose, and mouth).
The transform guides the GAN, showing which regions should be hallucinated via a convex hull $c(x)$ and which regions should be retained through a brain mask $b(x)$.
It also includes the brain data $b(x) \circ x$. 
The convex hull can afterwards be used to suppress noise patterns surrounding the head.
Following the preprocessing (see \emph{Appendix}), we define a function $c(x)$ that maps a scan $x \in \mathcal{I}^{S \times S \times S}$ to a binary convex hull volume of the same shape. 
As no efficient off-the-shelf algorithm exists, we developed a probabilistic solution that first constructs a surface representation from the MRI scan, and from this we compute the convex hull of the head. These steps are described below.

\begin{figure*}[t]
  \centering
  \includegraphics[width=0.95\linewidth]{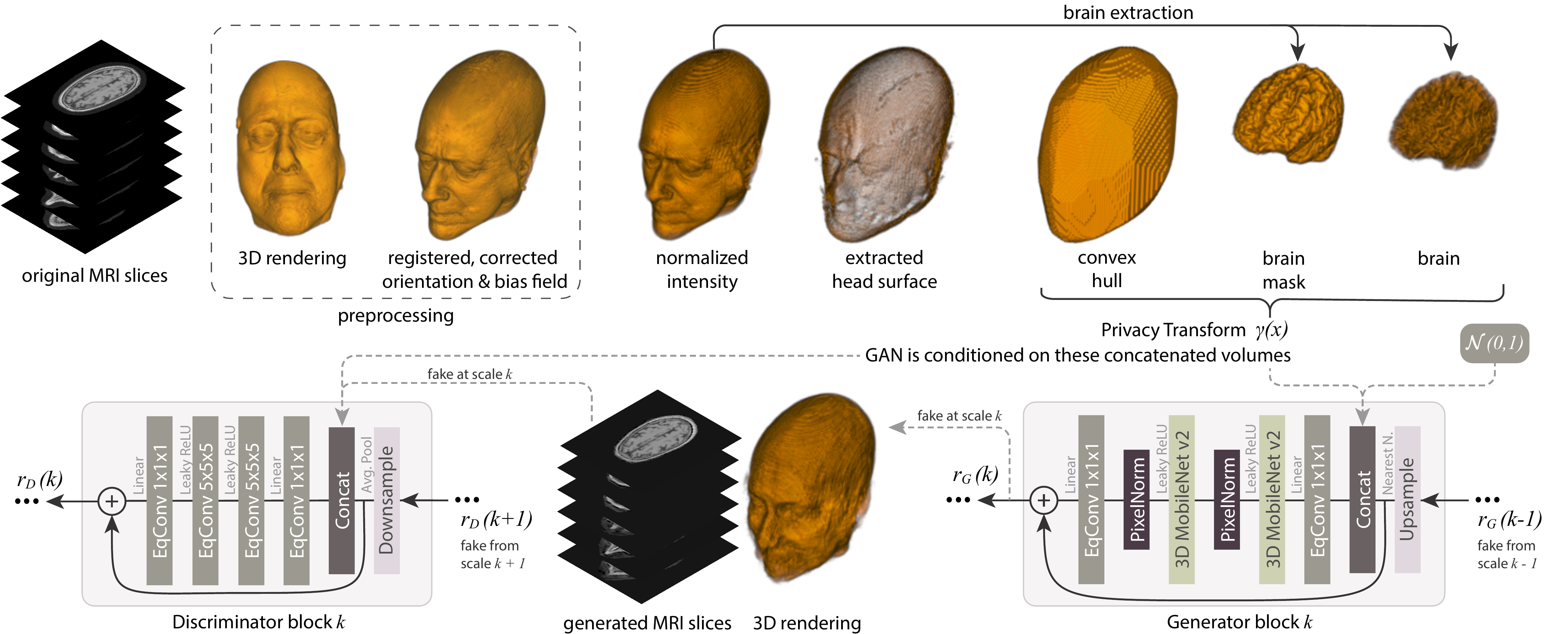}
  \vspace{3mm}
  \caption{\emph{Overview of our approach}: 
 We apply standard preprocessing to register and correct the scans (illustrated \dataset{OASIS-3}). Then, using a novel technique, we construct a \emph{head surface} representation from which we extract a convex hull $c(x)$, a brain mask $b(x)$, and the brain intensities $b(x) \circ x$. Combined, these form the privacy transform $\gamma(x)$ which serves as a conditioning variable of our model. \model{CP-GAN} learns to convert the distribution $\mathcal{P}_X$ of original MR scans to a de-identified counterpart $\mathcal{P}_Y$. Note that $\gamma(x)$ does not contain any privacy-sensitive information.  }
  \label{fig:overview}
  \vspace{-4mm}
\end{figure*}

\myparagraph{Surface Representation.}
To extract a surface representation $Z$ from an MRI scan $x$, we compute maps where rays cast from each direction intersect the head at random rotations. 
We then rotate these measurements back to the reference coordinates and treat 
each as the probability of it belonging to the surface. The rotations are randomized to sample the subject from all sides uniformly.
We begin by converting a given scan into a sequence of $K$ \emph{binarized} and \emph{rotated} scans, \ie $m^{(i)} = \text{Rot}(\mathds{1}[x \geq \delta]; R_i) \in \mathcal{I}^{S \times S \times S}$
for sampled rotations $R_1, \ldots, R_K \overset{\text{i.i.d}}{\sim} \mathcal{U}(\text{SO}(3))$, where $\mathcal{U}(\text{SO}(3))$ denotes the uniform distribution over all rotations in three-dimensional space and $\delta \in \mathcal{I}$ represents a suitably chosen binarization threshold\footnote{The threshold $\delta$ is chosen to be larger than the noise values surrounding the skull.} for the binarization operator $\mathds{1}[\cdot]$.
Let us further introduce the concept of the $\zeta_{a, d}$-distance of some voxel at position $(k_0, k_1, k_2)$ for some axis $a \in \lbrace 0, 1, 2 \rbrace$ and some direction $d \in \lbrace -1, +1 \rbrace$: 
\vspace{-3mm}
\begin{align}
    \zeta_{a, d}(k_0, k_1, k_2) \hat{=} \begin{cases}
                                                (S  - 1) - k_a & \text{if } d = +1 \\
                                                k_a & \, \text{otherwise.}
                                    \end{cases}
\end{align}
For fixed $a$ and $d$, we can use this to create an \emph{intersection map} $\Lambda_{a,d}^i$ for each binary image $m_S^i$:
\vspace{-4mm}
\begin{gather}
\begin{aligned}
    \Lambda_{a,d}^{(i)}[k_0, k_1, k_2] = \mathds{1} \bigg[ \bigg( m_{k_0,k_1,k_2}^{(i)} = 1  
    \bigg)
     \wedge
    \bigg( \zeta_{a,d}(k_0, k_1, k_2) =
    \mspace{-30mu} 
    \min_{\substack{s \in \lbrace 0, \ldots, S - 1\rbrace, \\ m_{k_{a-1} \mid  s \mid  k_{a+1}}^{(i)}  = 1}} 
    \mspace{-30mu} 
    \zeta_{a,d}(k_{a-1} \mid  s \mid  k_{a+1}) \bigg) \bigg] &&
 \end{aligned}
\raisetag{30pt}
\end{gather}
where $(k_{a-1} \mid  s \mid  k_{a+1}) $ indicates that the $a$-th index is set to $s$ and the two others to their associated value in $k_0, k_1, k_2$. 
We average the intersection map over all axis-direction combinations, \ie $\Lambda^{(i)} =\nicefrac{1}{6} \sum_a \sum_d \Lambda_{a,d}^i$.
This process can be thought of as casting rays from each principle direction and recording the location of the intersection with the rotated, binarized head in $m^{(i)}$.
Voxels on the surface of the head will exhibit high values of $\Lambda^{(i)}$. 
The final step is to back-rotate $\Lambda^{(1)}, \ldots, \Lambda^{(K)}$ to the reference coordinate system and average among the $K$ randomly sampled rotations to create the \emph{surface} representation:
\vspace{-3mm}
\begin{equation}
    Z = \nicefrac{1}{K}  \sum_{i=1}^K Rot(\Lambda^{(i)}; R_i^{-1}) \in [ 0, 1 ]^{S \times S \times S}
\end{equation}
Note that $Z$ is a random variable induced by the sampled rotations $R_1, \ldots, R_K$. We interpret individual voxel values of $Z$ as Bernoulli parameters characterizing the probability of some voxel belonging to the surface. This justifies binarizing $Z$ by considering it as a three-dimensional Bernoulli tensor and sampling from it on a voxel-wise basis in the next step.

\myparagraph{Convex Hull.}
From $Z$, we sample a set of non-zero indices and use Chan's Algorithm~\cite{Chan1996} to compute the triangles $\mathcal{T}$ making up the convex hull.
We initialize a uniform volume filled with 1's, then randomly select a sufficient number of triangles ($100$ suffice) from $\mathcal{T}$. 
For each triangle, we find its corresponding hyperplane and the half-spaces within $c(x)$ defined by it.
Voxels in the outward half-space of $c(x)$ are set to 0 while the rest are unchanged, yielding a binary convex hull volume.

\myparagraph{Privacy Transform.}
The binary convex hull volume $c(x)$ instructs the GAN as to which regions should be hallucinated.
A binary brain mask $b(x)$ obtained by applying~\cite{Iglesias2011} indicates which regions should be preserved.
Together, these volumes along with the masked
\emph{continuous} values of the brain $b(x) \circ x$, are concatenated to make the privacy transform $\gamma(x)$.
The GAN is conditioned on $\gamma(x)$  in the following subsection, as depicted in Figure \ref{fig:overview}.

\subsection{Conditional De-identification GAN}
The \model{CP-GAN} architecture depicted in Figure \ref{fig:overview} is capable of generating volumes at multiple scales and passing gradients between each scale during training. 
We start from a 2D generation framework akin to MSG-GAN~\cite{Karnewar2019} and adapt it to our task by means of the following: \emph{(1)} we incorporate \emph{conditional} information via  the privacy transform,
\emph{(2)} we make architectural improvements described below, \emph{(3)} we use a new resampling strategy, \emph{(4)} we adopt  relativistic (non-averaging) R-LSGAN loss, and \emph{(5)} we operate on 3D volumes.
We use \emph{bottlenecks} between scales as recently suggested by~\cite{Karras2019}, in which the generator outputs single-channel maps instead of multi-channel maps.
To reduce the memory footprint, we use modified MobileNetV2 convolutions as suggested in~\cite{DBLP:journals/corr/HowardZCKWWAA17}.

Both the generator $G_\Phi(\gamma(x))$ and the discriminator $D_\Theta(\gamma(x), v)$ are conditioned on $\gamma(x)$, where $v$ either denotes a multi-resolutional original or fake sample.
Regarding scales -- suppose that $S$ and $s$ are powers of two that denote the maximum/minimum resolution synthesized by $G_\Phi$. Then both $G_\Phi$ and $D_\Theta$ are defined to have $N_B = \log_2(\nicefrac{S}{s})  + 1$ blocks (indexed by $k$) that either double ($G_\Phi$) or halve ($D_\Theta$) their input resolution. Here, we generate scales from $4{\times}4{\times}4$ to $128{\times}128{\times}128$. 

\myparagraph{Generator.}
The generator $G_\Phi = G_{\Phi}^{(N_B)} \circ \ldots \circ G_{\Phi}^{(1)}$ for $G_{\Phi}^{(k)}: \mathbb{R}^{r_G(k-1)} \times \mathbb{R}^{r_G(k)} \rightarrow \mathbb{R}^{r_G(k)}$ and $ r_G(k) = 1 \times  2^{k-1} s \times 2^{k-1} s \times 2^{k-1} s$ synthesizes a sequence of fake images $g_1, \ldots, g_{N_B}$ of increasing resolutions as $g_k = G_{\Phi}^{(k)}(g_{k-1}, \gamma_k)\ \text{for}\ k=1,\ldots, N_B$ 
where $g_0 \sim \mathcal{N}(0,1) $ and $\gamma_k = \downdownarrows_{r_G(k)} \gamma(x)$ is $\gamma(x)$ downsampled to a resolution of $r_G(k)$.

\myparagraph{Discriminator.}
The discriminator $D_\Theta = F \circ D_{\Theta}^{(N_B)} \circ \ldots \circ D_{\Theta}^{(1)}$ for $D_{\Theta}^{(1)}:  \mathbb{R}^{r_D(1)} \times \mathbb{R}^{r_D(1)} \rightarrow \mathbb{R}^{r_D(1)}$ resp.\ $ D_{\Theta}^{(k)}: \mathbb{R}^{r_D(k-1)} \times \mathbb{R}^{r_D(k)} \times \mathbb{R}^{r_D(k)} \rightarrow \mathbb{R}^{r_D(k)} (k > 1)$ and $r_D(k) = 1 \times \nicefrac{S}{2^{k-1}} \times \nicefrac{S}{2^{k-1}} \times \nicefrac{S}{2^{k-1}}$ assigns a \emph{scalar} to a sequence of images\footnote{$x_1, \ldots, x_{N_B}$ in case of an original image and $g_1, \ldots, g_{N_B}$ in case of a fake image} of decreasing resolutions $v_1, \ldots, v_{N_B}$ as $d_1 = D_\Theta(v_1, \gamma_1)$ resp.\ $d_k = D_\Theta(d_{k-1}, v_k, \gamma_k)$ for $k = 2, \ldots, N_B$ 
where  $\gamma_k = \downdownarrows_{r_D(k)} \gamma(x)$ is $\gamma(x)$ downsampled to a resolution of $r_D(k)$ and $F$ is a fully-connected layer that computes a scalar summary of the output of $D_\Theta^{(N_B)}$.

\myparagraph{Resampling blocks.}
\cite{Karras2018b, Karras2019} recently proposed to use bilinear interpolation for downsampling, but adapting this approach is problematic as it will create undesirable interpolation effects in the binary volumes.
Therefore, we suggest a \emph{probabilistic} interpretation of \emph{average pooling}  which guarantees that the proportion of non-zero voxels is preserved (in expectation) while maintaining voxel-wise correspondence to conventional average pooling performed on non-binary images. Specifics can be found in the \emph{Appendix}.

\myparagraph{Loss Function.} We use the relativistic (\emph{non-averaging}) \emph{R-LSGAN} loss~\cite{Jolicoeur-Martineau2018}:
We opt for relativistic losses as they induce a lower memory footprint than, for instance, the widely-established WGAN-GP~\cite{gulrajani2017improved} requiring an additional forward/backward pass.

\myparagraph{Brain Preservation.}
One of the requirements defined above in the Problem Definition is to perfectly preserve medically relevant information.
Therefore, in a similar process to image inpainting in which original image content is masked and retained, we use the brain mask $b(x)$ to embed the original brain data into the volume synthesized by the generator, \ie $f_\Phi(\gamma(x)) = b(x) \circ x + (1 - b(x)) \circ G_\Phi(\gamma(x))$ where $\circ$ denotes the \emph{Hadamard} product.

\begin{figure*}[t]
\centering
\resizebox{\textwidth}{!}{%
    \begin{tabular}[t]{@{}c@{\hspace{2mm}}c@{\hspace{0.5mm}}c@{\hspace{0.5mm}}c@{\hspace{0.5mm}}c@{\hspace{4mm}}c@{\hspace{2mm}}c@{\hspace{0.5mm}}c@{\hspace{0.5mm}}c@{\hspace{0.5mm}}c@{}}

    {\footnotesize Original} & \multicolumn{4}{c}{\footnotesize \hspace{-3mm} Renderings after de-identification} &
    {\footnotesize Original} & \multicolumn{4}{c}{\footnotesize MRI slices after de-identification}\\
    \includegraphics[height=18mm]{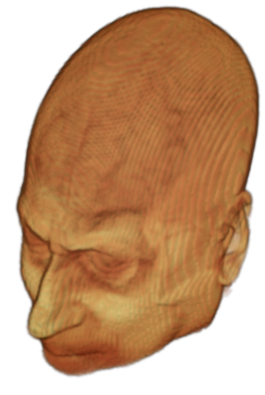}  \vline\hfill&
    \includegraphics[height=18mm]{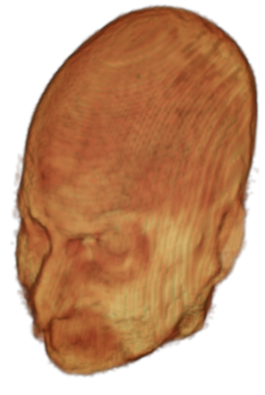} &
    \includegraphics[height=18mm]{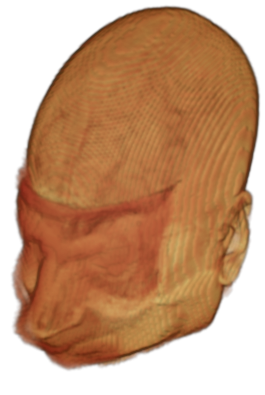} &
    \includegraphics[height=18mm]{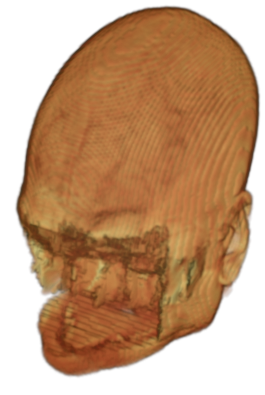} &
    \includegraphics[height=18mm]{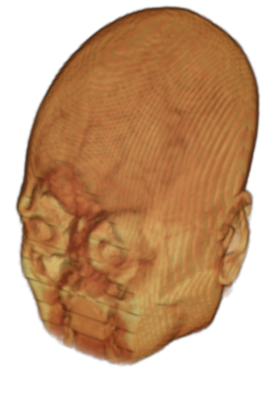}  &
    \includegraphics[height=19mm]{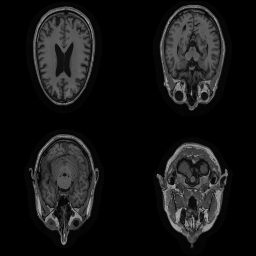}  &
    \includegraphics[height=19mm]{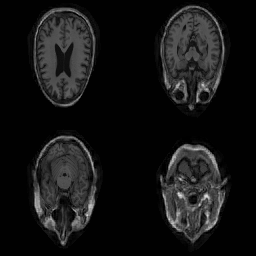}  &
    \includegraphics[height=19mm]{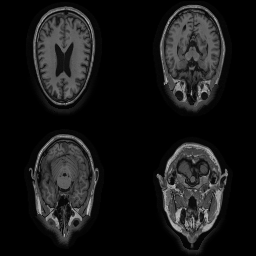}  &
    \includegraphics[height=19mm]{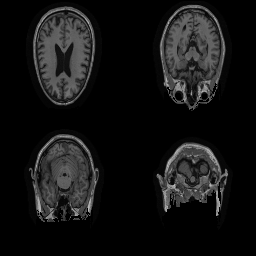}  &
    \includegraphics[height=19mm]{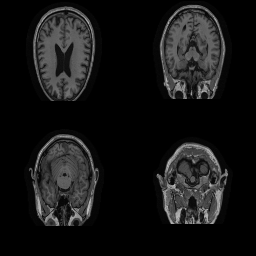}  \\[0mm]
    \includegraphics[height=18mm]{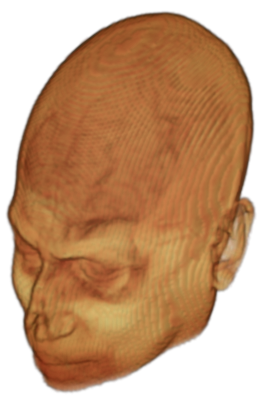}  \vline\hfill&
    \includegraphics[height=18mm]{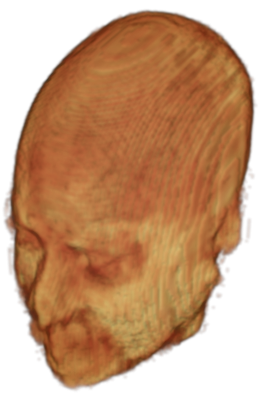} &
    \includegraphics[height=18mm]{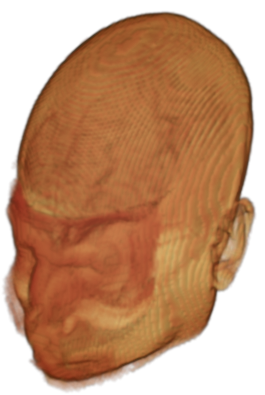} &
    \includegraphics[height=18mm]{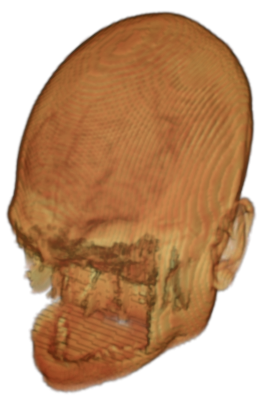} &
    \includegraphics[height=18mm]{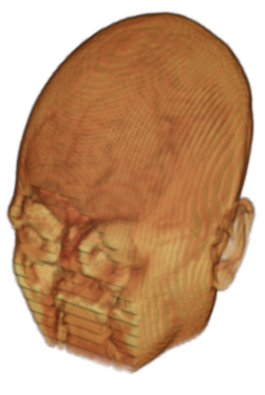}  &
    \includegraphics[height=19mm]{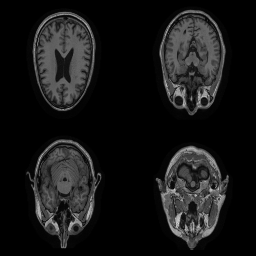}  &
    \includegraphics[height=19mm]{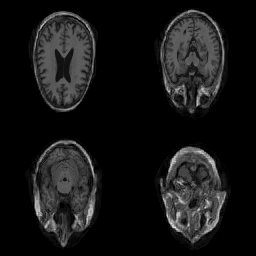}  &
    \includegraphics[height=19mm]{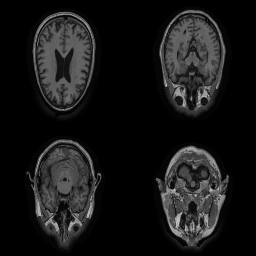}  &
    \includegraphics[height=19mm]{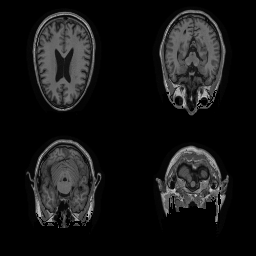}  &
    \includegraphics[height=19mm]{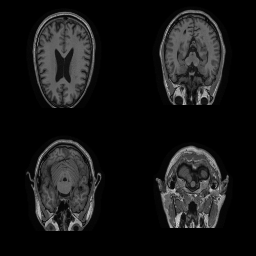}  \\[0mm]
    \includegraphics[height=18mm]{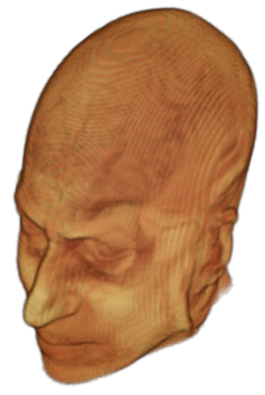}  \vline\hfill&
    \includegraphics[height=18mm]{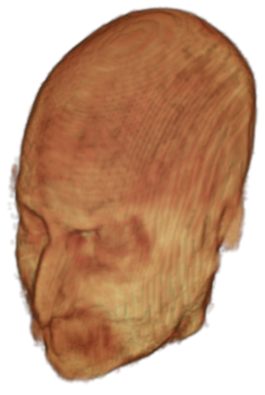} &
    \includegraphics[height=18mm]{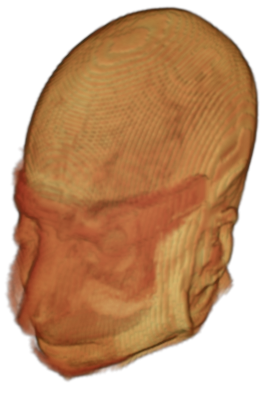} &
    \includegraphics[height=18mm]{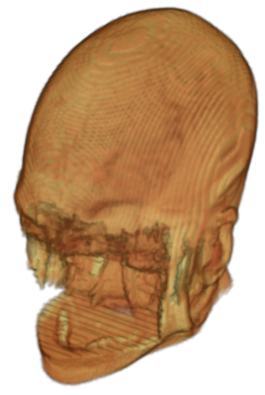} &
    \includegraphics[height=18mm]{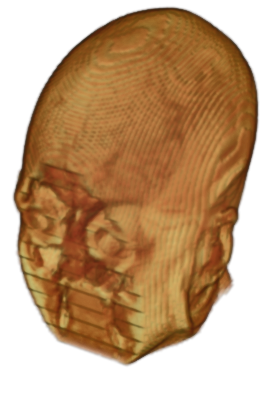}  &
    \includegraphics[height=19mm]{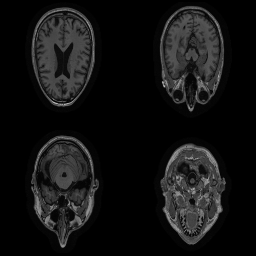}  &
    \includegraphics[height=19mm]{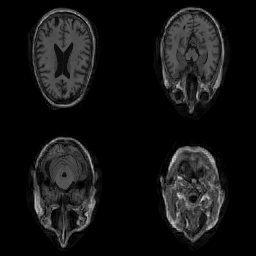}  &
    \includegraphics[height=19mm]{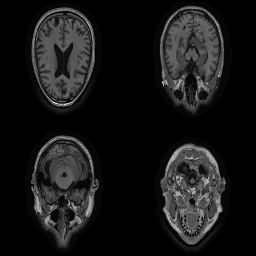}  &
    \includegraphics[height=19mm]{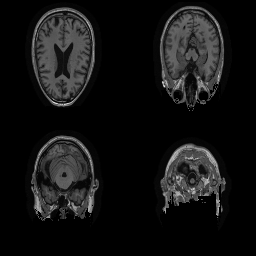}  &
    \includegraphics[height=19mm]{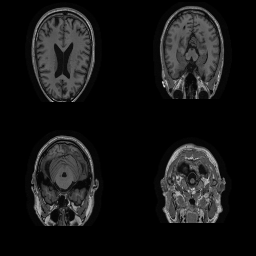}  \\[-1mm]
        & {\footnotesize \model{CP-GAN}} & {\footnotesize \model{FACE MASK}} & {\footnotesize \model{DEFACE}} & {\footnotesize \model{QUICKSHEAR}} &
    & {\footnotesize \model{CP-GAN}} & {\footnotesize \model{FACE MASK}} & {\footnotesize \model{DEFACE}} & {\footnotesize \model{QUICKSHEAR}}\\[-1mm]
\end{tabular}
}%
    \vspace{3mm}
    \caption{\emph{De-identification examples}: De-identified face renderings \emph{(left)} and the corresponding MRI scans \emph{(right)} for several examples from \dataset{OASIS-3}. All methods perfectly preserve the brain, but only \model{CP-GAN} successfully de-identifies the patient while maintaining realistic appearance and structure.}
    \label{fig:qualitative results}
    \vspace{-4mm}
\end{figure*}

\section{Experiments}
\label{sec:experiments}

Above, we proposed a new and modern approach to de-identify medical image data.
To judge its utility, we must address the following questions:
\emph{(1) Does remodeling preserve privacy better than existing removal-based de-identification methods?
(2) Does our approach adversely affect the performance of common medical applications?}
Below, we compare our approach to other de-identification methods to answer these questions experimentally.

\subsection{Setup}

\myparagraph{Datasets.}
In this work, we use two standard, publicly available large-scale Alzheimer's disease imaging studies which feature T1-weighted volumetric MR scans of the head for each subject: A selection of 2,172 MRIs from \dataset{ADNI}~\cite{Weiner2017,Wyman2013} and 2,168 MRIs from \dataset{OASIS-3}~\cite{LaMontagne2019}.
Both datasets are split ($80\%$-$20\%$ train-test) on a patient level to avoid data leakage by memorizing the patient.
The \dataset{ADNI} data used in the preparation of this article were obtained from the Alzheimer’s Disease
Neuroimaging Initiative database (\url{adni.loni.usc.edu}). \dataset{ADNI} was launched in
2003 as a public-private partnership, led by Principal Investigator Michael W. Weiner,
MD. The primary goal of \dataset{ADNI} has been to test whether serial magnetic resonance imaging
(MRI), positron emission tomography (PET), other biological markers, and clinical and
neuropsychological assessment can be combined to measure the progression of mild
cognitive impairment (MCI) and early Alzheimer’s disease (AD). For up-to-date information,
see \url{www.adni-info.org}.
Scanner types and acquisition protocols differ between and within the datasets, details can be found in the \emph{Appendix}.

\myparagraph{Benchmark De-Identification Methods.}
We compare our result with three publicly available and widely-established methods for de-identification of MRI head scans, depicted in Figure \ref{fig:qualitative results}. 
All methods have in common that they \emph{(1)} are not deep-learning-driven, \emph{(2)} require no additional training and \emph{(3)}, are used on a day-to-day basis in neuroscience and clinical research. 
All procedures were applied with default settings on images of resolution $128 {\times} 128 {\times} 128$. 
The methods include \model{QUICKSHEAR}~\cite{Schimke2011}, \model{FACE MASK}~\cite{Milchenko2013}, and \model{DEFACE}~\cite{Bischoff-Grethe2007}. 
Descriptions of the methods are provided in the \emph{Appendix}.
We also include \model{MRI WATERSHED}~\cite{Segonne2004}, a skull-stripping method that removes everything except the brain.

\myparagraph{Training.}
We use the AdamP~\cite{heo2020slowing} optimizer with a learning rate of $2 \cdot 10^{-3}$ and $\beta = (0, 0.99)$ and a batch size of 2. See the \emph{Appendix} for a complete list of hyperparameters.

\begin{figure}[t]
\begin{minipage}{0.5\textwidth}
    \resizebox*{\textwidth}{!}{
\begin{tabular}{lr@{\hspace{0.5mm}}lr@{\hspace{0.5mm}}lr@{\hspace{0.5mm}}lr@{\hspace{0.5mm}}l}
\toprule
\multicolumn{1}{c}{ } & \multicolumn{4}{c}{\tableHeading{USER-BASED}} & \multicolumn{4}{c}{\tableHeading{MODEL-BASED}}\\
\cmidrule(l{3pt}r{3pt}){2-5} \cmidrule(l{3pt}r{3pt}){6-9}
\multicolumn{1}{c}{ } & \multicolumn{2}{c}{\dataset{OASIS-3}} & \multicolumn{2}{c}{\dataset{ADNI}} & \multicolumn{2}{c}{\dataset{OASIS-3}} & \multicolumn{2}{c}{\dataset{ADNI}}\\
\cmidrule(l{3pt}r{3pt}){2-3} \cmidrule(l{3pt}r{3pt}){4-5} \cmidrule(l{3pt}r{3pt}){6-7} \cmidrule(l{3pt}r{3pt}){8-9}
\model{ORIGINAL}      &  $64.3$           & $\pm 2.1$          & $63.2$           & $\pm 2.1$           &  $100.0$          & $\pm 0.0$          & $100.0$             & $\pm 0.0$             \\ 
\model{BLURRED}       &  $57.2$           & $\pm 2.1$          & $53.1$           & $\pm 2.2$          &  $46.7$          & $\pm 8.2$          & $51.2$           & $\pm 7.8$             \\ \hdashline
\model{FACE MASK}     &  $48.7$           & $\pm 2.3$          & $47.5$          & $\pm 2.2$          &  $68.5$           & $\pm 7.6$          & $63.3$           & $\pm 8.2$             \\
\model{DEFACE}        &  $54.0$           & $\pm 2.3$          & $46.7$          & $\pm 2.3$          &  $70.7$          & $\pm 6.4$          & $73.1$            & $\pm 7.0$             \\
\model{QUICKSHEAR}    &  $45.6$           & $\pm 2.3$          & $39.9$          & $\pm 2.2$          &  $44.0$           & $\pm 7.4$          & $47.0$           & $\pm 6.7$             \\
\model{CP-GAN}    &  $ \mathbf{28.3}$ & $\mathbf{\pm 2.0}$ & $\mathbf{22.4}$ & $\mathbf{\pm 1.8}$ &  $\mathbf{29.5}$  & $\mathbf{\pm 5.9}$ & $\mathbf{27.4}$  & $\mathbf{\pm 5.7}$    \\ \hdashline
\model{BLACK}         &  $18.9$           & $\pm 1.6$          & $20.5$          & $\pm 1.8$          &  $19.7$          & $\pm 4.7$          & $19.8$           & $\pm 4.4$             \\
\model{MRI WATERSHED} &  $20.7$           & $\pm 2.0$          & $18.7$          & $\pm 1.6$          &  $21.0$          & $\pm 4.8$          & $21.6$           & $\pm 4.4$             \\
\bottomrule
\end{tabular}
    }
\end{minipage}%
\ 
\begin{minipage}{0.5\textwidth}
    \begin{minipage}{0.46\textwidth}
        \centering{ {\scriptsize \fontfamily{phv} {OASIS-3}}}\\
         \includegraphics[height=3cm]{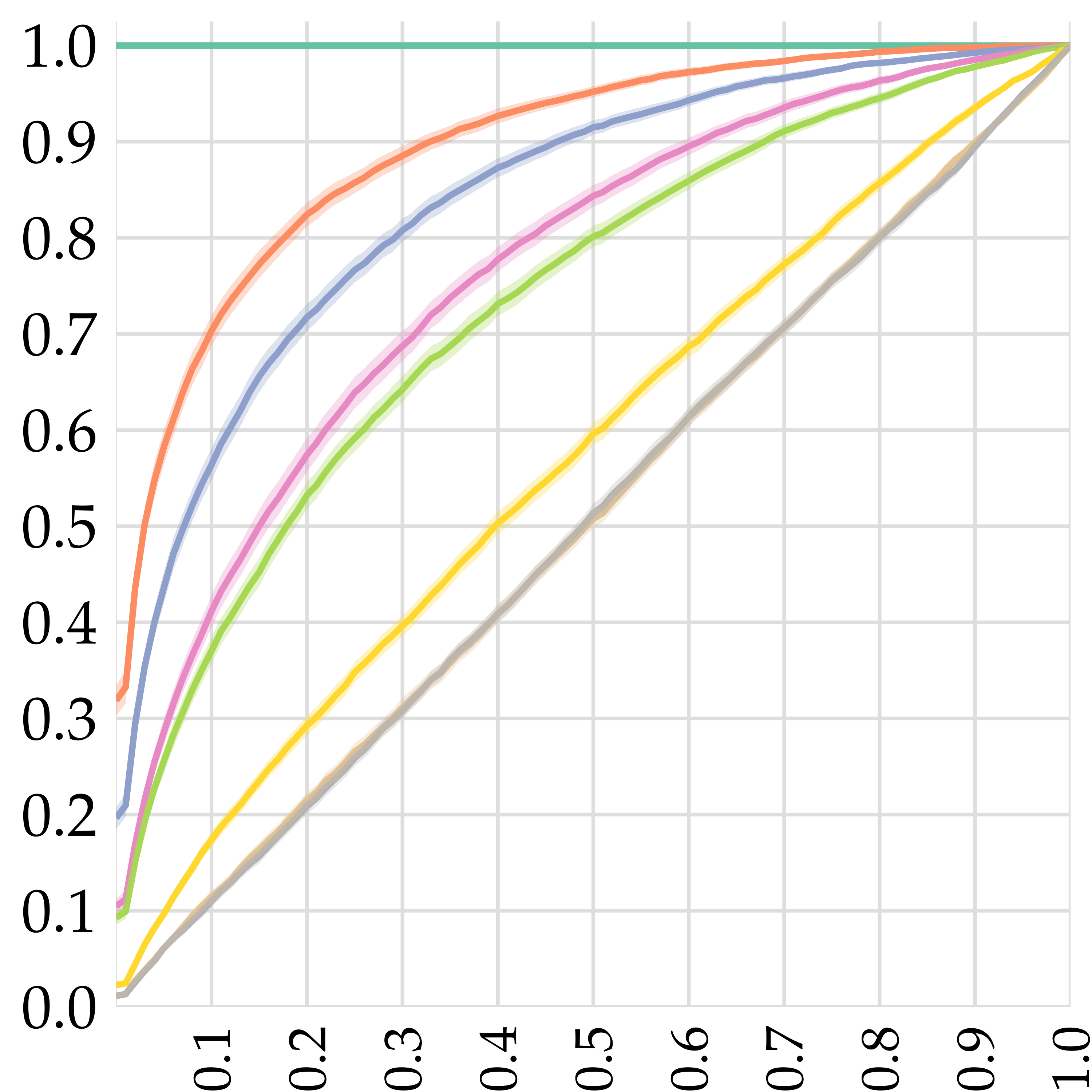}\\
         \vspace{-1.5mm}
         {\scriptsize Relative rank $\alpha$}
    \end{minipage}
    \ 
    \begin{minipage}{0.46\textwidth}
         \centering{{\scriptsize \fontfamily{phv} {ADNI}}}\\
         \includegraphics[height=3cm]{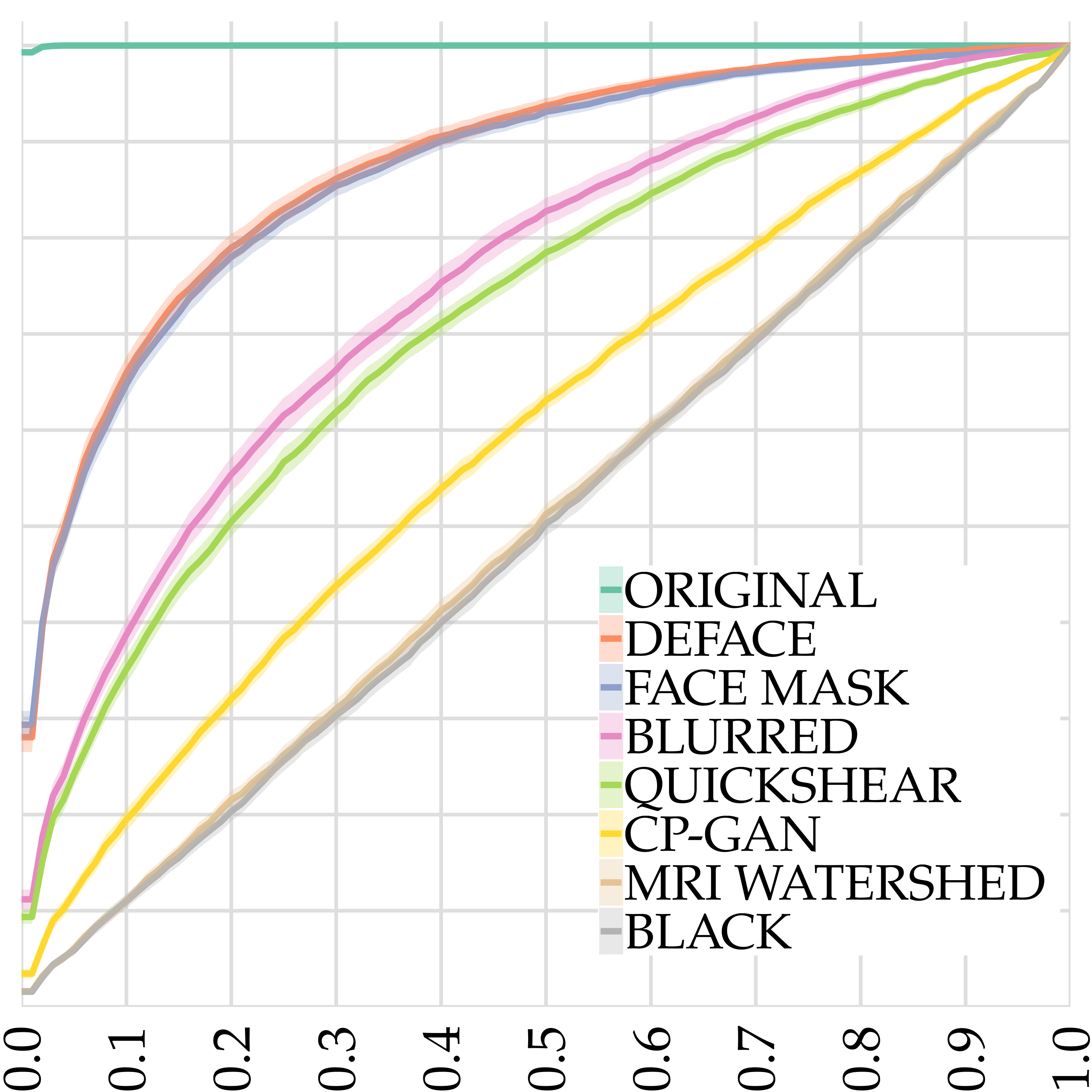}\\
         \vspace{-1.5mm}
         {\scriptsize Relative rank $\alpha$}
    \end{minipage}%
    \begin{minipage}{0.03\textwidth}
    \vspace{-3mm}
    {\scriptsize
    \begin{turn}{270}
         $\mathrm{Pr}[\mathcal{C} \in \mathrm{Top}(\alpha)]\ \ {\scriptstyle (\hat{=} \mathrm{Pr}[\alpha_{\mathcal{C}} \leq \alpha])}$
    \end{turn}
    }
    \end{minipage}%
\end{minipage}%
\vspace*{2mm}
     \caption{(\emph{left}) A user-based study on AMT and a model-based study with \emph{Siamese} networks to determine how well de-identified faces can be identified. Correct identification rates ($\pm$s.d.)\ are reported for two datasets (lower is better); (\emph{right}) A Siamese network takes a face rendering as an input, and aims to identify the correct de-identified rendering in a retrieval scenario. We measure the probability of the correct face appearing in its top $\alpha$-percentile ranked predictions. \model{BLACK} = \emph{optimal}, \model{ORIGINAL} = \emph{pessimal}. }
     \label{fig:deidentification_results}
\end{figure}

\subsection{Results}
In this section, we present results on (1) studies comparing the identification rate of our model with existing de-identification methods, and (2) the effects of de-identification on common medical image analysis tasks.
In the \emph{Appendix}, we provide a comparison of execution times. Video results are provided in \emph{supplementary material.}

\myparagraph{De-identification quality user study.}
The privacy attack described in~\cite{mazura2012facial} relied on prospectively collected data, meaning the authors had access to CT scans as well as photographs of patient faces.
Replicating that study for MRI scans is impossible, because photographs of
\dataset{ADNI} and \dataset{OASIS-3} patients do not exist.
Therefore, we conduct a similarly-spirited study using \emph{Amazon Mechanical Turk} in which workers are asked to defeat the various de-identification methods given renderings of MRI scans.
Workers were presented with an unaltered rendering of a query patient along with five renderings de-identified using a single method\footnote{An exemplary question can be found in the \emph{Appendix}} -- one of which is a de-identified rendering of the query patient. 
The task was then to pick out the de-identified rendering which corresponds to the unaltered query rendering.
We considered the following de-identification methods: \model{QUICKSHEAR}~\cite{Schimke2011}, \model{FACE MASK}~\cite{Milchenko2013}, and \model{DEFACE}~\cite{Bischoff-Grethe2007}, and \model{CP-GAN} (ours). 
In addition to the four de-identification methods, we added four control tasks, \model{ORIGINAL}, which signifies the absence of any de-identification scheme, \model{BLURRED}, in which the 2D renderings are blurred to mildly obscure the patient identity, \model{BLACK}, which features the same all-black image for each option, and \model{MRI WATERSHED}~\cite{Segonne2004} which completely removes all tissue except the brain.
We asked 800 distinct questions per dataset. 
Each question was given to five workers, for a total of 4,000 assignments. 
The mean and the standard deviation are estimated by bootstrapping over 1,000 resamples.

In Figure \ref{fig:deidentification_results} (\emph{left}), we report the identification rate, or how often the workers were able to defeat each method, see \emph{Appendix} for details.
The upper performance bound from random guessing corresponds to 20\%.
The results substantiate the claim that \model{CP-GAN} performs extraordinarily well at de-identification.
Our model outperforms the other de-identification methods by gaps of 17\%--25\% on both datasets. 
We note that for both datasets, \model{CP-GAN} performs close to the theoretical optimum of 20\%.

\myparagraph{De-identification quality model-based study.}
In a similar fashion to the last experiment, we assess the de-identification performance of the various models by attempting to defeat them.
This time, however, we leverage a neural network to assess similarity.

To this end, we use a metric learning approach to train a Siamese network $\mathcal{S}(\cdot, \cdot)$ to quantify whether its two input renderings belong to the same patient or not.

\begin{table}
\centering
    \resizebox*{\textwidth}{!}{
    \begin{tabular}{@{}l@{\hspace{3mm}}c@{\hspace{3mm}}c@{\hspace{3mm}}c@{\hspace{3mm}}c@{\hspace{3mm}}c@{\hspace{3mm}}c@{\hspace{3mm}}c@{\hspace{3mm}}c@{\hspace{3mm}}c@{\hspace{3mm}}c@{\hspace{3mm}}c@{\hspace{3mm}}c@{\hspace{3mm}}c@{\hspace{3mm}}c@{\hspace{3mm}}c@{\hspace{3mm}}c@{}}
        \toprule
        \multicolumn{1}{c}{ } & \multicolumn{8}{c}{\emph{Sørensen-Dice coefficient $\uparrow$}} & \multicolumn{8}{c}{\emph{Intersection-over-Union} (IoU) $\uparrow$} \\
        \cmidrule(lr{3pt}){2-9} \cmidrule(lr{3pt}){10-17}
        \multicolumn{1}{c}{ } & \multicolumn{4}{c}{\tableHeading{OASIS-3}}  & \multicolumn{4}{c}{\tableHeading{ADNI}} & \multicolumn{4}{c}{\tableHeading{OASIS-3}}  & \multicolumn{4}{c}{\tableHeading{ADNI}}\\
        \cmidrule(lr{3pt}){2-5} \cmidrule(lr{3pt}){6-9} \cmidrule(lr{3pt}){10-13} \cmidrule(lr{3pt}){14-17} 
        & BRAIN & VCSF & WHITE & GREY & BRAIN & VCSF & WHITE & GREY & BRAIN & VCSF & WHITE & GREY & BRAIN & VCSF & WHITE & GREY \\
        \model{ORIGINAL} & 1.000 & 1.000 & 1.000 & 1.000 & 1.000 & 1.000 & 1.000 & 1.000  & 1.000 & 1.000 & 1.000 & 1.000 & 1.000 & 1.000 & 1.000 & 1.000    \\
         \hdashline[2pt/1pt]
        \model{FACE MASK} & 0.991 & 0.984 & 0.989 & 0.996 & 0.986 & 0.977 & 0.976 & 0.987 & 0.982 & 0.968 & 0.978 & 0.992 & 0.973 & 0.955 & 0.953 & 0.975 \\
        \model{DEFACE} & 0.993 & 0.986 & 0.986 & 0.995 & 0.982 & 0.966 & 0.965 & 0.981 & 0.985 & 0.972 & 0.973 & 0.990 & 0.965 & 0.934 & 0.932 & 0.963 \\
        \model{QUICKSHEAR} & 0.994 & 0.989 & 0.990 & 0.997 & 0.986 & 0.975 & 0.972 & 0.985 & 0.987 & 0.978 & 0.980 & 0.994 & 0.972 & 0.952 & 0.946 & 0.971 \\
        \model{CP-GAN} & \textbf{0.995} & \textbf{0.991} & \textbf{0.992} & \textbf{0.998} & \textbf{0.989} & \textbf{0.979} & \textbf{0.978} & \textbf{0.989} & \textbf{0.989} & \textbf{0.981} & \textbf{0.983} & \textbf{0.996} & \textbf{0.977} & \textbf{0.960} & \textbf{0.957} & \textbf{0.978} \\
         \hdashline[2pt/1pt]
        \model{MRI WATERSHED} & 0.675 & 0.415 & 0.564 & 0.718 & 0.717 & 0.570 & 0.589 & 0.732 & 0.509 & 0.262 & 0.393 & 0.560 & 0.559 & 0.399 & 0.417 & 0.578 \\
        \bottomrule
    \end{tabular}
    }
\vspace{2.5mm}
\caption{\emph{Brain segmentation:} We measure the Sørensen-Dice coefficient and the IoU computed between segmentations on the original scan and de-identified scans using standard software, SIENAX. 
We test on the whole brain, VCSF, white matter, and grey matter.
Ideally, segmentation should not be affected by de-identification, indicated by a Dice score and IoU of 1.
\model{CP-GAN} outperforms all other de-identification methods.
Note that \model{MRI WATERSHED}, which removes everything but the brain, has a catastrophic effect. 
}
\label{fig:medical_tasks}
\end{table}
Given two inputs $x$ and $y$, the network $\mathcal{S}$ is constructed by applying a sub-network $\tilde{\mathcal{S}}$ (conv.\ block/flatten/fully-connected layer) on $x$ and $y$ independently, followed by summarizing both embeddings with the \emph{Euclidean} distance, \ie $\mathcal{S}(x, y) = \lVert \tilde{\mathcal{S}}(x) - \tilde{\mathcal{S}}(y) \rVert_2 $. We use the \emph{Triplet Margin} loss function as described in~\cite{BMVC2016_119}, choosing the margin to be equal to $5$. We split the previously defined (hold-out) data set, and \emph{randomly} select 80\% of its patients as a training set $\mathcal{D}_i$ and the remaining 20\% as its complement $\overline{\mathcal{D}_i}$, where $i=1,\ldots,100$ denotes the $i$-th (resampled) fold. Specifics on the nature of the training can be found in the \emph{Appendix}. 

In Figure \ref{fig:deidentification_results} (\emph{left}) we report the ability of the network to defeat the de-identification methods in similar fashion to the \emph{user-based} study.
We first sample a patient $p$ from $\overline{\mathcal{D}}_{i}$ from whom, in turn, we sample a scan $s \in S(p)$. Afterwards, we sample a method $m$ and and consider the $m$-rendering of $s$ to be the \emph{correct} option $\mathcal{C}$. 
The remaining $5-1=4$ options are obtained by randomly selecting $m$-renderings from other patients' scans. Denoting the five options by $x_{1}, \ldots, x_{5}$, we obtain the predicted option by $k = \argmin_{j=1,\ldots,5} \mathcal{S}(x_{O}, x_{j})$ where $x_{O}$ denotes the original rendering of $s$.
As in the \emph{user-based} study, \model{CP-GAN} outperforms the other de-identification methods.
For \model{FACE MASK} and \model{DEFACE}, the network was able to de-identify between 16 to 20\% more renderings than its human counterparts.
The effect on \model{QUICKSHEAR} is more moderate.

In Figure \ref{fig:deidentification_results} (\emph{right}) we evaluate the Siamese network's ability to defeat de-identification methods in a retrieval-inspired setting.
We observe that \model{CP-GAN}'s de-identification capabilities are strikingly close to the \emph{optimal} case while other methods perform substantially worse. 
For some given original rendering $x^{(\textrm{orig.})}$ and some method $m$, we analyze how often the correct choice $\mathcal{C}$ falls within the top $\alpha$ (relative) ranks, \eg $\mathrm{Top}(\alpha{=}0.1)$ is a subset of the $10\%$ top-ranked scans. 
An \emph{optimal} de-identification method induces a \emph{uniformly random} rank $\alpha_{\mathcal{C}} \sim \mathcal{U}(0, 1)$ of $\mathcal{C}$ (c.f.\ \model{BLACK}), whereas a \emph{pessimal} method induces a \emph{Dirac} placement $\alpha_{\mathcal{C}} = 0$ (c.f.\ \model{ORIGINAL}). Confidence bands ($\textrm{CI}{=}0.95$) are calculated over the $100$ previously defined data splits $\overline{\mathcal{D}}_i$.

Overall, we conclude that \model{CP-GAN} outperforms established methods by a substantial double-digit margin, withstanding both human and model-based attacks.
We note a gap between \model{CP-GAN} and optimal performance, which can be explained by a property of the model: it preserves head size.
The attacker may exploit this to eliminate some candidates.
We explore this concept further in the \emph{Appendix}.

\myparagraph{Effect of De-Identification on Medical Analyses.} 
Beyond ensuring patient privacy, de-identification methods should not adversely affect software tools commonly used on medical scans.
However, it has been shown that facial de-identification methods \emph{do} adversely impact automated image analysis on MRI scans used in research and in the clinic~\cite{DeSitter2020}.  In line with this study, we conduct two experiments. In the first, we assess how the de facto standard brain tissue segmentation tool, SIENAX~\cite{Smith2004}, performs on de-identified MRI scans in comparison to the originals.
In Table \ref{fig:medical_tasks}, we report the (Sørensen-) Dice scores~\cite{sorensen1948method,Dice1945} and IoU between the original and de-identified scans for various brain segmentation tasks. 
We observe that \model{CP-GAN} outperforms all of its contenders, proving that brain volume estimations are reliable after the subject is de-identified using  \model{CP-GAN}. 
\emph{Note that removing everything except the brain using} \model{MRI WATERSHED} \emph{has a catastrophic effect}, replicating the effect observed in~\cite{Fennema}.

In the second experiment, we investigate whether de-identification adversely affects brain age estimation -- an important task as the difference between predicted and chronological age has links to  brain disease~\cite{jonsson2019brain}. 
This is a challenging task for \model{CP-GAN} since age information captured in the MRI is filtered out in $\gamma(x)$ in contrast to the other methods that preserve head information in addition to the brain. 
Nonetheless, we find that our de-identification introduces less bias than \model{DEFACE} and \model{QUICKSHEAR} on both datasets, though \model{FACE MASK} slightly outperforms our model.  Due to space limitations, these results appear in the \emph{Appendix}.

\vspace{-5mm}

\section{Conclusion}
In this work, we defined a new paradigm for de-identification of medical imagery and realized it for MRI scans.
Our approach 
remodels privacy-relevant information while keeping medically-relevant information untouched. 
\emph{It can be applied to other modalities, producing remodeled images that appear genuine and preserve relevant medical information, but without revealing privacy-sensitive information.} 
Our method protects privacy substantially better than existing methods, without compromising analyses typically found in research and clinical settings -- a crucial deficiency of strong removal methods such as skull-stripping. 
A future research direction is to extend our approach to other MRI and CT modalities, adding new downstream tasks such as lesion and brain tumor segmentation~\cite{SCHMIDT20123774,meierraphael}. Incorporating other pulse sequences, such as \emph{T2-weighting} or \emph{FLAIR}, requires a re-training of the network as well as a sufficient number of training samples. Unfortunately, alternative pulse sequences are less readily available in comparison to the \emph{T1-weighted} imagery used in this paper. Apart from this apparent scarcity, we however do not expect any change in complexity.
We hope that the methods outlined here can help to better protect patient privacy.

\vspace{-6mm}
\paragraph{Acknowledgements}
\footnotesize
\textit{
This work was partially supported by the Swedish Research Council (VR) 2017-04609, the ERC (853489
- DEXIM) and by the DFG (2064/1 – Project number 390727645).
Data collection and sharing for this project was funded by the Alzheimer's Disease
Neuroimaging Initiative (ADNI) (National Institutes of Health Grant U01 AG024904) and
DOD ADNI (Department of Defense award number W81XWH-12-2-0012). ADNI is funded
by the National Institute on Aging, the National Institute of Biomedical Imaging and
Bioengineering, and through generous contributions from the following: AbbVie, Alzheimer’s
Association; Alzheimer’s Drug Discovery Foundation; Araclon Biotech; BioClinica, Inc.;
Biogen; Bristol-Myers Squibb Company; CereSpir, Inc.; Cogstate; Eisai Inc.; Elan
Pharmaceuticals, Inc.; Eli Lilly and Company; EuroImmun; F. Hoffmann-La Roche Ltd and
its affiliated company Genentech, Inc.; Fujirebio; GE Healthcare; IXICO Ltd.; Janssen
Alzheimer Immunotherapy Research \& Development, LLC.; Johnson \& Johnson
Pharmaceutical Research \& Development LLC.; Lumosity; Lundbeck; Merck \& Co., Inc.;
Meso Scale Diagnostics, LLC.; NeuroRx Research; Neurotrack Technologies; Novartis
Pharmaceuticals Corporation; Pfizer Inc.; Piramal Imaging; Servier; Takeda Pharmaceutical
Company; and Transition Therapeutics. The Canadian Institutes of Health Research is
providing funds to support ADNI clinical sites in Canada. Private sector contributions are
facilitated by the Foundation for the National Institutes of Health (www.fnih.org). The grantee
organization is the Northern California Institute for Research and Education, and the study is
coordinated by the Alzheimer’s Therapeutic Research Institute at the University of Southern
California. ADNI data are disseminated by the Laboratory for Neuro Imaging at the
University of Southern California.}
\normalsize
\bibliography{bibliography}
\end{document}